\RequirePackage[hyphens]{url}
\documentclass[10pt]{iopart}
\expandafter\let\csname equation*\endcsname=\relax
\expandafter\let\csname endequation*\endcsname=\relax
\usepackage{graphicx,amsmath,amssymb,epsfig,paralist,tikz,pgfplots}
\pgfplotsset{compat=newest}

\renewcommand{\vec}[1]{\boldsymbol{\mathrm{#1}}}
\allowdisplaybreaks
\begin{document}

\title{Pushing limits: Probing new gravity using a satellite constellation}

\author{Viktor T. Toth$^1$\footnote{Corresponding author ({\tt vttoth$@$vttoth.com})}}

\address{$^1$Ottawa, Ontario K1N 9H5, Canada}

\date{\today}

\begin{abstract}

Building upon earlier work, we explore the limits of using a configuration of satellites to measure the trace of the gravitational gradient tensor using intersatellite laser ranging and timing observables without relying on high-precision external observables such as deep space radio navigation or astrometry with unrealistic accuracy. A refined model, calculated with extended numerical precision, confirms that exceptional sensitivity is possible, placing within reach observational tests of certain modified gravity theories (e.g., Yukawa terms, galileons) using heliocentric orbits in the vicinity of the Earth. The sensitivity of the experiment improves at larger heliocentric distances. A constellation placed at 30 astronomical units, still well within the domain of feasibility using available propulsion and deep space communication technologies, may approach sensitivities that are sufficient to detect not just the gravitational contribution of the interplanetary medium but perhaps even cosmological dark matter and dark energy constituents.

\end{abstract}



\section{Introduction}

Recently, we investigated the use of a tetrahedral satellite configuration in heliocentric orbit, as a tool to measure the trace of the gravitational gradient tensor \cite{TOTH2023ApSS}. A practical measurement of this type, if carried out at sufficient accuracy, may offer a means of direct detection of a dark perfect fluid background \cite{Yu2019} or the presence of terms characteristic of modified gravity theories (e.g., galileon field \cite{Nicolis2009}).

A critical aspect of this conceptual experiment is that it relies entirely (or almost entirely) on intersatellite observables, and does not require an external reference with unrealistic expectations on astrometry or radio-metric navigation, far beyond what is technically feasible. It is possible in principle, utilizing only intersatellite laser ranging and precise timing observables, to reconstruct the shape of the tetrahedral configuration at very high fidelity, as well as the rate of rotation of a reference frame that is attached to this configuration. By derotating the reference frame and establishing the satellites' inertial acceleration, the trace of the gravitational gradient tensor becomes accessible. The experiment can be carried out without radio navigation or astrometry at unrealistic levels of accuracy.

In \cite{TOTH2023ApSS}, we presented a detailed description of this system, along with a computational model, which was implemented in the form of a software simulation, used also in \cite{Turyshev2024}. We also presented an analysis of modeling errors, concluding that, in principle, a tetrahedral configuration of satellites may be able to detect ${\cal O}(3\times 10^{-23}~{\rm s}^{-2})$ or smaller deviations from zero in the trace of the gravitational gradient tensor when orbiting the Sun at $\sim 1$~AU (astronomical unit). Orbits with larger semimajor axes allow for even greater sensitivity.

The main limitation of this software simulation was due to the constraints on accuracy due to double-precision arithmetic, hindering our ability to model sub-micron ranging observables in a heliocentric context. To address this concern, we have now endeavored to replicate our results using an extended precision implementation of the same computation. As we shall see below, once this model is used in combination with an improved numerical integrator, numerical differences between precomputed (in an inertial frame) and modeled (in the constellation-affixed rotating frame) observables vanish, and any deviation from a zero trace of the gravitational gradient tensor that remains can be attributed to fundamental limitations inherent to the approach being considered (notably that the gravitational gradient is reconstructed over finite intersatellite distances, not infinitesimal line elements, and that the rate of rotation of the satellite-fixed reference frame is measured to first order only).

These results lend confidence to the notion that the trace of the gravitational gradient tensor may prove to be a tool to explore non-Newtonian contributions to gravity, such as those proposed by many popular modified gravity theories. With our simulation tool, we can not only confirm theoretical predictions but also explore possible configurations that may, in the future, be used to measure (or constrain) contributions from such modified gravity theories within the solar system. Although the technological challenges may prove formidable, a project may benefit from past or on-going efforts, such as high accuracy intersatellite laser ranging experiment of the GRACE follow-on mission \cite{REL2014} or the inertial sensor assemblies of LISA/LISA Pathfinder \cite{LISA2013}.

We begin in Section~\ref{sec:recap} by briefly recapping the relationship between the constellation dynamics and the trace of the gravitational gradient tensor. In Section~\ref{sec:model}, we present the revised numerical model and our new results, extended to orbits out to 30~AU, which show that remarkable observational accuracy may be achievable in principle. We put these results to use in Section~\ref{sec:postnewton}, by modeling three popular modifications of Newtonian gravity: a Yukawa force, a galileonic field and Modified Newtonian Dynamics; we also discuss the consequences of working within a Newtonian approximation. In Section~\ref{sec:summary} we summarize these results and offer conclusions.

\section{The gravitational gradient tensor and the constellation}
\label{sec:recap}

We follow the derivation first presented in \cite{TOTH2023ApSS}, which we briefly recap here. Our starting point is Newtonian gravity.

\subsection{Newtonian gravity}

As it is well known, the gravitational acceleration $\vec{a}$ in a Newtonian gravitational potential $U$ is determined by the gradient of the potential:
\begin{align}
\vec{a}=-\nabla U.
\end{align}
For a compact (point) mass, its gravitational potential at a distance $r$ from that source is of course $U=-GM/r$. This expression is the Green's function solution of the general field equation that characterizes Newtonian gravity, Poisson's equation for gravitation:
\begin{align}
\nabla^2 U=4\pi G\rho,
\end{align}
where $G$ is Newton's constant for gravitation and  $\rho$ is the mass density characterizing the distribution of matter.

The left-hand side of Poisson's equation can also be written in terms of $\vec{a}$:
\begin{align}
\nabla^2 U = \nabla\cdot(\nabla U)=-\nabla\cdot\vec{a}.
\end{align}

This quantity can also be viewed as the trace of the gravitational gradient tensor, which is defined as
\begin{align}
\vec{T}=-\nabla\otimes\nabla U,
\end{align}
where $\otimes$ is the outer product. Using this definition, we can write Poisson's equation in the form
\begin{align}
-\tr\vec{T}=\nabla^2 U=4\pi G\rho.
\end{align}

In the vacuum, the gravitational potential satisfies $\nabla^2 U=0$. That is to say, the trace of the gravitational gradient tensor vanishes in the vacuum.

\begin{figure}
\begin{center}
\begin{tikzpicture}
\draw[color=gray!25,line width=0.1pt] (-0.5cm,0.5cm) -- (4cm,2cm);
\draw[color=gray!25,line width=0.1pt] (-1cm,1cm) -- (3.5cm,2.5cm);
\draw[color=gray!25,line width=0.1pt] (-1.5cm,1.5cm) -- (3cm,3cm);
\draw[color=gray!25,line width=0.1pt] (-2cm,2cm) -- (2.5cm,3.5cm);

\draw[color=gray!25,line width=0.1pt] (0.75cm,0.25cm) -- (-1.75cm,2.75cm);
\draw[color=gray!25,line width=0.1pt] (1.5cm,0.5cm) -- (-1cm,3cm);
\draw[color=gray!25,line width=0.1pt] (2.25cm,0.75cm) -- (-0.25cm,3.25cm);
\draw[color=gray!25,line width=0.1pt] (3cm,1cm) -- (0.5cm,3.5cm);
\draw[color=gray!25,line width=0.1pt] (3.75cm,1.25cm) -- (1.25cm,3.75cm);

\draw[color=red!20,line width=0.5pt] (0,0) -- (-1cm,2cm);
\draw[color=red!20,line width=0.5pt] (0,0) -- (1.5cm,5cm);
\draw[color=red!20,line width=0.5pt] (3cm,1cm) -- (-1cm,2cm);
\draw[color=red!20,line width=0.5pt] (3cm,1cm) -- (1.5cm,5cm);
\draw[color=red!20,line width=0.5pt] (-1cm,2cm) -- (1.5cm,5cm);
\draw[->] (0,0) -> (4.5cm,1.5cm);
\draw[->] (0,0) -> (-2.5cm,2.5cm);
\draw[->] (0,0) -> (0cm,6cm);
\filldraw[color=green] (3cm,1cm) circle (0.1cm);
\filldraw[color=blue] (-1cm,2cm) circle (0.1cm);
\filldraw[color=orange] (1.5cm,5cm) circle (0.1cm);
\draw[->,line width=0.75pt] (0,0) -> (3cm,1cm);
\draw[->,line width=0.75pt] (0,0) -> (-1cm,2cm);
\draw[->,line width=0.75pt] (0,0) -> (1.5cm,5cm);
\filldraw[color=red] (0,0) circle (0.1cm);
\node[anchor=west] at (4.5cm,1.5cm) {$x$};
\node[anchor=east] at (-2.5cm,2.5cm) {$y$};
\node[anchor=south] at (0cm,6cm) {$z$};
\node[anchor=north west] at (0,0) {$l$};
\node[anchor=north west] at (3cm,1cm) {$\vec{r}_{li}$};
\node[anchor=south east] at (-1cm,2cm) {$\vec{r}_{lj}$};
\node[anchor=south west] at (1.5cm,5cm) {$\vec{r}_{lk}$};
\end{tikzpicture}
\end{center}
\caption{\label{fig:coords}The four-satellite configuration and coordinate system.}
\end{figure}
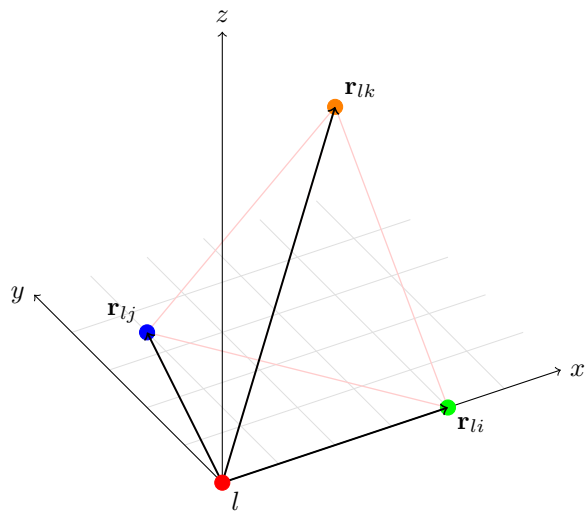

This statement can be tested experimentally by a measurement involving a set of four free-flying satellites (Fig.~\ref{fig:coords}). The gravitational acceleration of a test particle (or any object suitable as a test particle, such as a free-flying satellite) with trajectory $\vec{r}=\vec{r}(t)$ is given by $\ddot{\vec{r}}=\vec{a}=-\nabla U$. The gravitational gradient tensor may be given in terms of the acceleration field as $\vec{T}=\nabla\otimes\vec{a}$, or alternatively, written in terms of infinitesimals,
\begin{align}
\vec{T}\cdot d\vec{r}=d\vec{a}.
\end{align}
Replacing infinitesimal displacements with finite relative position vectors $\vec{r}_{ij}$ between satellite $i$ and $j$, we obtain
\begin{align}
\vec{T}\cdot\vec{r}_{ij}=\vec{a}_{ij}
\end{align}
Arranging a set of three non-coplanar vectors $\vec{r}_{li}$, $\vec{r}_{lj}$, $\vec{r}_{lk}$ into a matrix $\vec{P}$ and the corresponding accelerations into $\vec{A}$, we obtain
\begin{align}
\vec{T}\cdot\vec{P}=\vec{A},\label{eq:TP=A}
\end{align}
or
\begin{align}
\vec{T}=\vec{A}\cdot\vec{P}^{-1},
\end{align}
with
\begin{align}
\vec{P}^{-1}=\frac{1}{\vec{r}_{li}\cdot(\vec{r}_{lj}\times\vec{r}_{lk})}\begin{bmatrix}{}[\vec{r}_{lj}\times\vec{r}_{lk}]^T\\{}[\vec{r}_{lk}\times\vec{r}_{li}]^T\\{}[\vec{r}_{li}\times\vec{r}_{lj}]^T\end{bmatrix},
\end{align}
which leads to a simple formula for the trace:
\begin{align}
\tr\vec{T}=\sum\limits_{ijk}\frac{\vec{a}_{li}\cdot[\vec{r}_{lj}\times\vec{r}_{lk}]}{\vec{r}_{li}\cdot[\vec{r}_{lj}\times\vec{r}_{lk}]},
\label{eq:trT}
\end{align}
with the summation carried out over even permutations of the index triplet $ijk$.

The promise of this form is that if the relative positions and accelerations between a set of four satellites can be measured, the trace of the gravitational field tensor can be recovered.

\subsection{The satellite-fixed reference frame}

To put this formulation to work, first the relative position vectors $\vec{r}_{ij}$ must be recovered. Given a configuration of four satellites, six intersatellite range measurements allow us to recover the relative position vectors $\vec{r}_{ij}$ in a satellite-fixed reference frame. The calculations are somewhat tedious but algebraically trivial. For the sake of completeness, let us recap the key results (previously published in this form in \cite{TOTH2023ApSS}.) See also Fig.~\ref{fig:coords} for reference.

First, we define a set of basis vectors $\vec{e}_{x..z}$:
\begin{align}
\vec{e}_x & {} = \frac{\vec{r}_{li}}{|\vec{r}_{li}|},\\
\vec{e}_z & {} = \frac{\vec{r}_{li}\times\vec{r}_{lj}}{|\vec{r}_{li}\times\vec{r}_{lj}|}\\
\vec{e}_y & {} = \vec{e}_z\times\vec{e}_x.
\end{align}
We then express the position vectors $\vec{r}_{li..k}$ corresponding to the four satellites in this satellite-fixed reference frame. 
\begin{align}
\vec{r}_{li}&{}=[r_{li},0,0],\\
\vec{r}_{lj}&{}=\left[\frac{r_{li}^2 + r_{lj}^2 - r_{ij}^2}{2r_{li}}, \sqrt{r_{lj}^2 - \left(\frac{r_{li}^2 + r_{lj}^2 - r_{ij}^2}{2r_{li}}\right)^2}, 0\right],\\
\vec{r}_{lk}&{}=\begin{bmatrix}\dfrac{r_{li}^2 + r_{lk}^2 - r_{ik}^2}{2r_{li}},\\~\\
 \dfrac{r_{lj}^2 + r_{lk}^2 - r_{jk}^2 - \dfrac{(r_{li}^2 + r_{lk}^2 - r_{ik}^2)(r_{li}^2 + r_{lj}^2 - r_{ij}^2)}{2r_{li}^2}}{2\sqrt{r_{lj}^2 - \left(\dfrac{r_{li}^2 + r_{lj}^2 - r_{ij}^2}{2r_{li}}\right)^2}},\\~\\
 \pm\sqrt{r_{lk}^2 - \left(\dfrac{r_{li}^2 + r_{lk}^2 - r_{ik}^2}{2r_{li}}\right)^2 - \left(\dfrac{r_{lj}^2 + r_{lk}^2 - r_{jk}^2 - \dfrac{(r_{li}^2 + r_{lk}^2 - r_{ik}^2)(r_{li}^2 + r_{lj}^2 - r_{ij}^2)}{2r_{li}^2}}{2\sqrt{r_{lj}^2 - \left(\dfrac{r_{li}^2 + r_{lj}^2 - r_{ij}^2}{2r_{li}}\right)^2}}\right)^2}\end{bmatrix}.\label{eq:pos}
\end{align}

\subsection{Accelerations}

The reference frame in which we recovered the components of $\vec{r}_{ij}$ is not inertial. Using a time series of $\vec{r}_{ij}$ and finite differences, we can recover a corresponding $\vec{a}_{ij}$ but it will include fictitious contributions due to the rotation of the reference frame. To remove these contributions, the rate of rotation of the reference frame must be established.

This is actually possible using only intersatellite observables, by utilizing a Sagnac-type \cite{Sagnac1,Sagnac2} measurement: differences in round-trip times along any three of the satellites in the clockwise vs. counterclockwise direction. Choosing, in particular, three faces of the tetrahedron adjoining a vertex, if these three faces are non-coplanar (i.e., if the tetrahedral configuration is non-degenerate with zero volume), the three observables can be used to recover, to first order, the angular velocity vector $\vec{\omega}$. of the rotating reference frame with the origin at the chosen vertex. The governing equations are \cite{TOTH2023ApSS}:
\begin{align}
ct_{ki}&{}=|\vec{r}_{ki}+t_{ki}(\vec{v}_{ki}-\vec{\omega}\times\vec{r}_{ki})|,\label{eq:SagnacB1}\\
ct_{ij}&{}=|\vec{r}_{kj}+(t_{ki}+t_{ij})(\vec{v}_{kj}-\vec{\omega}\times\vec{r}_{kj})-[\vec{r}_{ki}+t_{ki}(\vec{v}_{ki}-\vec{\omega}\times\vec{r}_{ki})]|,\\
ct_{jk}&{}=|\vec{r}_{kj}+(t_{ki}+t_{ij})(\vec{v}_{kj}-\vec{\omega}\times\vec{r}_{kj})|.\label{eq:SagnacB3}
\end{align}
In the inertial reference frame, when $\vec{\omega}=0$, these equations allow us to construct the value of the Sagnac-type observable,
\begin{align}
\Delta t_{kij}=(t_{ki}+t_{ij}+t_{jk}) - (t_{kj}+t_{ji}+t_{ik}).\label{eq:Delta_t}
\end{align}
In the rotating reference frame of the satellite constellation, $\vec{\omega}\ne 0$, positions and velocities must be replaced with the corresponding primed quantities measured in that reference frame, which correspond to observables. Thereafter, we can solve for the value of $\vec{\omega}$ that yields the correct ``observed'' value of the Sagnac-type observable:
\begin{align}
\Delta t_{kij}=(t_{ki}'+t_{ij}'+t_{jk}') - (t_{kj}'+t_{ji}'+t_{ik}').\label{eq:DeltaSagnac}
\end{align}
Performing this exercise on the three faces of the tetrahedron that adjoin a particular vertex, we can solve for the three independent degrees of freedom of $\vec{\omega}$.

Thereafter, we can employ Rodrigues' formula \cite{Rodrigues1840} for the rotation of a vector over a finite time interval $\Delta t$, $\vec{r}'=(\vec{I}+(\sin\theta)\vec{K}+(1-\cos\theta)\vec{K}^2)\cdot\vec{r}$, with the rotation matrix $\vec{K}=|\vec{\omega}|^{-1}\vec{\Omega}$, where the angular velocity tensor $\Omega=\star\omega$ is the Hodge-dual of the angular velocity vector and $\theta=|\vec{\omega}|\Delta t$. Utilizing the rotation matrix, we can establish a relationship between a rotating and a nonrotating reference frame that coincide at some moment in time $t$, with both of them having the same origin at a given vertex $l$. At any other time $t+\Delta t$, relative position vectors between the origin and other vertices in the rotating and nonrotating reference frames are related, to first order, by
\begin{align}
\vec{r}'_{li}(t+\Delta t)=(\vec{I}+\Delta t\vec{\Omega})\vec{r}_{li}(t+\Delta t).
\end{align}
Using this formula to re-express $\vec{r}'_{ij}$ (which are obtained from intersatellite ranging) as $\vec{r}_{ij}$ (in an inertial frame), we are finally in a position to calculate accelerations using numerical differentiation. Improving upon the standard formulation used in \cite{TOTH2023ApSS}, we can use a five-point stencil to numerically compute second derivatives using derotated relative position vectors:
\begin{align}
\vec{a}(t)\approx\frac{-\vec{K}^2\cdot\vec{r}'(t-2\Delta t) + 16\vec{K}\cdot\vec{r}'(t-\Delta t) - 30\vec{r}'(t) + 16\vec{K}^{-1}\cdot\vec{r}'(t+\Delta t) - \vec{K}^{-2}\vec{r}'(t+2\Delta t)}{12(\Delta t)^2}.
\label{eq:stencil}
\end{align}

Once we are in possession of $\vec{a}_{ij}$ in an inertial reference frame, we can compute (\ref{eq:trT}), as inner products and triple products are invariant under rotation.

Before concluding, however, we must take note of two additional concerns.

First, there exists a chiral ambiguity: a tetrahedron and its mirror image cannot be distinguished using only the six intersatellite ranges. We can resolve this ambiguity by ensuring that the handedness of the tetrahedron, with respect to the order in which we enumerate its vertices, corresponds to the handedness of the reference frame in which the coordinates are expressed.

Second, numerically calculating accelerations using the method outlined above does not take into account that when the baseline between satellites is large enough, the acceleration field can no longer be treated as constant along that baseline. Assuming that the gravitational field is dominated by the Sun, and that we know the {\em approximate} direction $\vec{n}$, and distance $r$, from the Sun, we can introduce the correction \cite{TOTH2023ApSS}:
\begin{align}
\delta\vec{a}_{ij}\approx-\frac{3GM}{r^4}\left[\frac{3}{2}|\vec{r}_{ij}|^2\vec{n}
-\frac{5}{2}(\vec{n}\cdot\vec{r}_{ij})^2\vec{n}
+\vec{r}_{ij}\times(\vec{r}_{ij}\times\vec{n})\right].
\end{align}
We emphasize that as this correction is small in magnitude, the Sun direction and heliocentric distance need not be known precisely in the satellite-fixed reference frame. Therefore, even though this quantity represents an observable that references an external object (the Sun), the required precision of the observation remains achievable. In our simulation, we deliberately introduced an error by using the approximate center of the tetrahedron, as opposed to the satellite that serves as the origin of the satellite-fixed reference frame. This represents an error of the same magnitude as the intersatellite ranges, i.e., typically several hundred km or more, and realistically corresponds to the accuracy with which the constellation's orbit may be known, e.g., from standard radio navigation.

\section{Numerical modeling and results}
\label{sec:model}

We have thus established that using a time series of intersatellite range observables, Sagnac-type observables along triplets of satellites, and the approximate distance and direction from the Sun, it is possible to recover the trace of the gravitational gradient tensor. This result can be used in the form of a numerical simulation with two distinct stages.

First, in a conveniently chosen (e.g., heliocentric) inertial  reference frame, the simulation progresses the orbits of the spacecraft and computes observables: specifically, intersatellite distances and the Sagnac-type observables (\ref{eq:Delta_t}) for each tetrahedron face, as discussed in the preceding section.

Next, ``firewalled'' from any preceding calculation that took place in the inertial reference frame, the simulation uses time series data sets for the six intersatellite ranges, the four Sagnac-type timing results, and the approximate distance and direction to the Sun, and {\em nothing else}. Using the intersatellite ranges, satellite coordinates are established in a satellite-fixed reference frame. This rotating reference frame is derotated using the Sagnac-type observables augmented by the approximate Sun direction. Relative accelerations are calculated using derotated coordinates, leading to the computation of the trace of the gravitational gradient tensor.

Separating the code that models the trajectories of the system of satellites from the code that models the measurement performed by these satellites was critical to our ability to analyze what such a satellite configuration can measure as a gravitational instrument in deep space.

The accuracy to which this computation can be carried out is limited primarily by our omission of higher order acceleration terms. The most stringent limitation on accuracy arises from our failure to account for angular accelerations. We estimated the magnitude of the error due to this \cite{TOTH2023ApSS}:
\begin{align}
\delta_{[\omega]}\tr{\vec{T}}=\left(\frac{d}{1,000~{\rm km}}\right)\left(\frac{1~{\rm AU}}{a}\right)^{4.5}{\cal O}(3\times 10^{-23}~{\rm s}^{-2}).
\label{eq:err}
\end{align}

As a realistic scenario, we considered orbits that were not ``perfect'' configurations. Initial positions and velocities were slightly perturbed, as it can be expected in any real-life experiment: Propulsion systems are not perfect, valves don't always close consistently, outgassing occurs, and other factors may be present that will limit our ability to control initial conditions. This justified the use of initial conditions that contained small {\em ad hoc} perturbations, as shown in Table~\ref{tab:incond}. We of course assume that once the constellation is in flight and the experiment commences, individual satellites are unaffected by any nongravitational influences, including on-board sources such as outgassing or anisotropic thermal radiation, as well as environmental sources such as solar radiation pressure or drag by the interplanetary medium. (Practical implementations may utilize, e.g., freely falling assemblies contained within a protective spacecraft enclosure with a controlled environment therein.)

\begin{table}
\caption{\label{tab:incond}Initial state vectors characterizing the constellations that were used in the simulations presented in Figs.~\ref{fig:str1AU}--\ref{fig:ranges}. Individual satellite positions and velocities are relative to the nominal constellation position and velocity. All nominal constellation state vectors were initialized to start on the $x$-axis, with the initial velocity in the $y$-direction. Eccentricities ($\epsilon$) and orbital periods ($T$) are also shown for each constellation.}~\par
\footnotesize{\begin{tabular}{l|c|c|c|c|c|c|c}
Case&S/C&$x$&$y$&$z$&$v_x$&$v_y$&$v_z$\\\hline\hline
1~AU&~&0.6~AU&~&~&~&48.5~km/s&~\\\cline{2-8}
$\epsilon=0.5909$&$i$&$-500$~km&$-500$~km&$-500$~km&$+0.1$~m/s&$+0.17$~m/s&$+0.5$~m/s\\
$T=648.80$~days&$j$&$+500$~km&$-500$~km&$+500$~km&$+0.1$~m/s&$-0.17$~m/s&$-0.8$~m/s\\
~&$k$&$-500$~km&$+500$~km&$+500$~km&$+0.1$~m/s&$+0.17$~m/s&$-0.4$~m/s\\
~&$l$&$+500$~km&$+500$~km&$-500$~km&$-0.1$~m/s&$-0.17$~m/s&$+0.3$~m/s\\\hline\hline
10~AU&~&10~AU&~&~&~&8~km/s&~\\\cline{2-8}
$\epsilon=0.2786$&$i$&$-1000$~km&$-1000$~km&$-3000$~km&$+0.001$~m/s&$~~~~~~0$~m/s&$~+0.01$~m/s\\
$T=7989.33$~days&$j$&$+1000$~km&$-1000$~km&$+2000$~km&$~~~~~~~0$~m/s&$-0.02$~m/s&$~~~~~~~0$~m/s\\
~&$k$&$-1000$~km&$+1000$~km&$+5000$~km&$~~~~~~~0$~m/s&$~~~~~~0$~m/s&$~+0.01$~m/s\\
~&$l$&$+1000$~km&$+2000$~km&$-2000$~km&$~~~~~~~0$~m/s&$-0.02$~m/s&$~~~~~~~0$~m/s\\\hline\hline
30~AU&~&30~AU&~&~&~&5~km/s&~\\\cline{2-8}
$\epsilon=0.1546$&$i$&$-1000$~km&$-1000$~km&$-3000$~km&$+0.001$~m/s&$~~~~~~~0$~m/s&$+0.0001$~m/s\\
$T=48377.88$~days&$j$&$+1000$~km&$-1000$~km&$+2000$~km&$~~~~~~~0$~m/s&$-0.002$~m/s&$~~~~~~~~0$~m/s\\
~&$k$&$-1000$~km&$+1000$~km&$+5000$~km&$~~~~~~~0$~m/s&$~~~~~~~0$~m/s&$+0.0001$~m/s\\
~&$l$&$+1000$~km&$+2000$~km&$-2000$~km&$~~~~~~~0$~m/s&$-0.002$~m/s&$~~~~~~~~0$~m/s\\\hline\hline
\end{tabular}}
\end{table}

\begin{figure}
\includegraphics[scale=0.8]{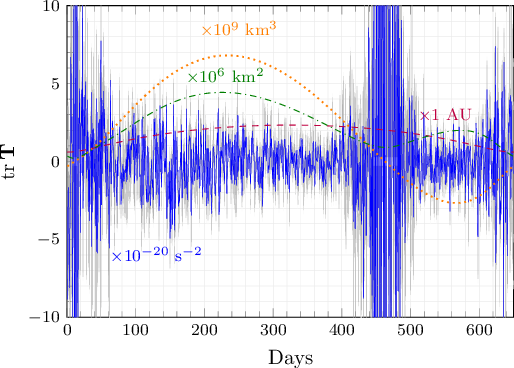}~\includegraphics[scale=0.8]{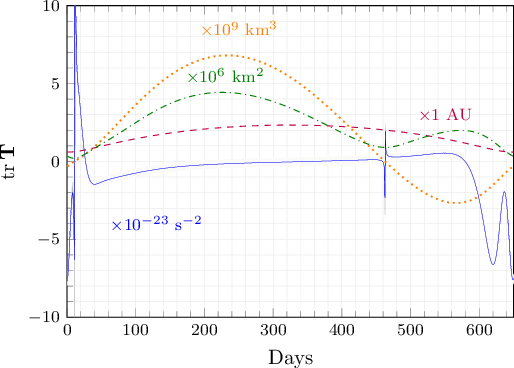}
\caption{\label{fig:str1AU}Simulated measurement of $\tr\vec{T}$ for a tetrahedral configuration in a $\sim 650$~day orbit with a perihelion of $\sim 0.6$~AU. Left: results calculated at standard machine precision (64-bit double precision, $\sim 15.9$~decimal digits.) Right: The exact same simulation, recalculated using 24 decimal digits. Blue (solid) line: $\tr\vec{T}$. (Note the difference in vertical scale between the two results.) Orange (dotted) line: the oriented volume of the tetrahedron. Green (dash-dotted) line: the tetrahedron surface area. Red (dashed) line: heliocentric distance.}
\end{figure}

When we attempted to validate the model's limits of accuracy through simulation, we encountered a hard technical limit. Standard double-precision representation of floating point quantities yields an accuracy of just over 15 decimal digits. Usually, this is more than sufficient for modeling physical systems, but not in this case. The difference in magnitude between heliocentric distances (of ${\cal O}(10^{11}~{\rm m})$ or greater) vs. the required intersatellite range accuracy to model, in particular, the Sagnac-type observables from which the satellite-fixed reference frame's angular velocity is derived, imposes a stringent limit. Assuming a cumulative relative error of $10^{-14}$ after chain calculations, we obtained the error estimate
\begin{align}
\delta_{[\omega]}^{\tt num}\tr\vec{T}=\left(\frac{1~{\rm AU}}{a}\right)^{1.5}\left(\frac{1,000~{\rm km}}{d}\right) {\cal O}(1.2\times 10^{-18}~{\rm s}^{-2}).
\end{align}
This is several orders of magnitude greater than the desired, anticipated sensitivity of the experiment using a $\sim 1$~AU orbit. Making a more optimistic assumption on the cumulative rounding error, while also employing some numerical ``tricks of the trade'' in the software code to mitigate errors (this could be something as simple as first computing sums or differences of small quantities before adding the result to a larger quantity), we can reduce this numerical error further, but it remains substantial, as demonstrated, e.g., by Fig.~\ref{fig:str1AU} (left).

This numerical error can be addressed definitively, however, by carrying out the simulation using extended precision software, albeit at the cost of an increase in software complexity and a substantial performance penalty. We endeavored to carry out this exercise as the best means to validate the accuracy of our model and our estimates concerning its inherent capabilities and limitations.

For this reason, we reimplemented our numerical model of the satellite configuration, using the {\tt Decimal.js} extended precision numerical library \cite{decimaljs}. We expected a significant slowdown as a result of using a software floating point implementation with extended precision, but nonetheless, it was important to explore the capabilities of this representation of the satellite configuration, without being constrained by machine precision limitations. We also anticipated that calculating the model using extended precision might reveal any subtle errors or shortcomings that were not apparent previously, being masked by the numerical noise due to standard double-precision arithmetic\footnote{Indeed, we found a subtle bug that was present in the earlier version of our software, used in \cite{TOTH2023ApSS} and also \cite{Turyshev2024}: the code used to resolve the chiral ambiguity of the reconstructed tetrahedron contained an error, yielding a contribution comparable in magnitude to, and therefore masked by, the noise due to the limited accuracy of the double precision representation.}.

In the process, we also endeavored to update the numerical model. To improve the accuracy of numerical derivatives, calculated using the method of finite differences from time series of intersatellite ranges, without the need to use an unreasonably low time step that would slow down the computation, we opted to use the five-point stencil (\ref{eq:stencil}) mentioned in the previous section. Furthermore, the original code used a standard fourth-order Runge-Kutta integrator \cite{Runge1895,Kutta1901}, but we now also experimented with a fourth-order Yoshida integrator \cite{Yoshida1990} as well as an implementation of St{\"o}rmer's rule \cite{stormer1907} in conjunction with the Richardson \cite{Richardson1911} approximation, which we used previously in navigational code \cite{TOTH2008}. Both of these choices were motivated, in part, to employ an integrator that is consistent with energy conservation. In the end, however, the three integrators proved equally effective and accurate when used with small timesteps. Our current results, in the form of the plots shown here, were computed using the St{\"o}rmer's rule integrator, which we chose based on its symplectic nature, known accuracy and efficiency, and history of success. Source code for the updated implementation is available on GitHub\footnote{\url{https://github.com/vttoth/TETRA}}.

Our first result is presented in Fig.~\ref{fig:str1AU}, modeling the $\sim 1$~AU orbit of a four-satellite constellation, initially forming a regular tetrahedron with a volume of $3.333\times 10^8~{\rm km}^3$. As the plots indicate, once the limitation due to double precision arithmetic is addressed, the computation yields a very clean result, with a magnitude that is consistent with our error estimate (\ref{eq:err}). As a matter of fact, any significant difference between the $\tr\vec{T}$ computed in the inertial reference frame of the simulation, vs. the simulated measurement of $\tr\vec{T}$ that was  reconstructed from intersatellite ranges and Sagnac-type observables vanishes. This computation demonstrates explicitly that even for an orbit in the vicinity of the Earth, measuring $\tr\vec{T}$ with an accuracy of $\sim 10^{-23}~{\rm s}^{-2}$ is achievable in principle, consistent with our analytical estimates in \cite{TOTH2023ApSS}.

\begin{figure}
\includegraphics[scale=0.8]{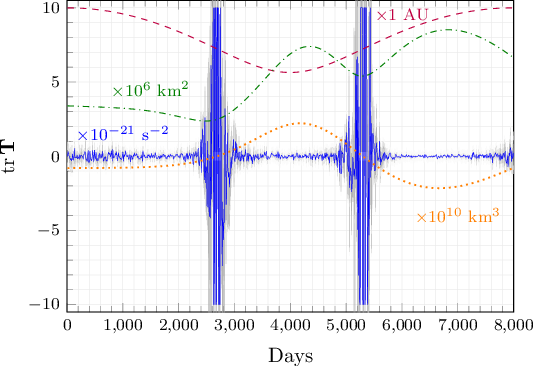}~\includegraphics[scale=0.8]{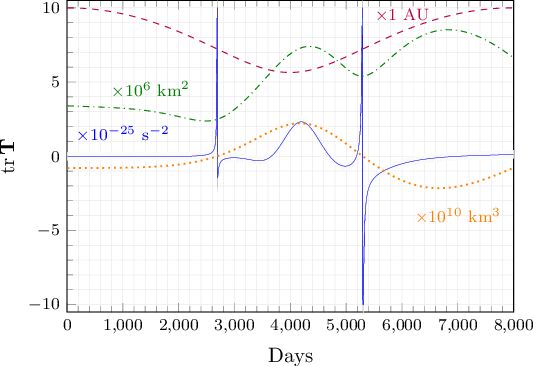}
\caption{\label{fig:str10AU}Simulation of a $\sim 10$~AU orbit. Comparison of standard double precision (left) vs. extended precision (28 digits, right) shows that once numerical artifacts are eliminated, the tetrahedral configuration shows remarkable sensitivity throughout an entire orbit, except for brief transitional periods when the tetrahedron becomes degenerate, its volume collapsing. Colors and line types are the same as in Fig.~\ref{fig:str1AU}.}
\end{figure}

Encouraged by this result, we also performed a simulation of a much larger, $\sim 10$~AU orbit with a similar tetrahedral configuration. The result, shown in Fig.~\ref{fig:str10AU}, reveals that much greater sensitivity is possible in the outer solar system.

\begin{figure}
\includegraphics[scale=0.8]{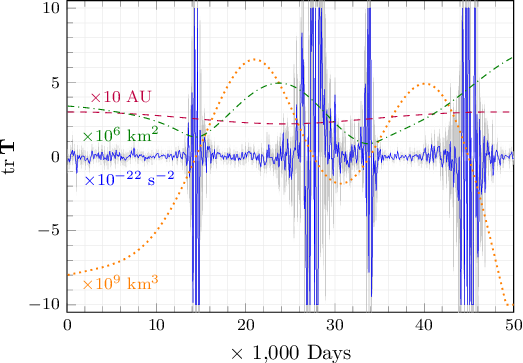}~\includegraphics[scale=0.8]{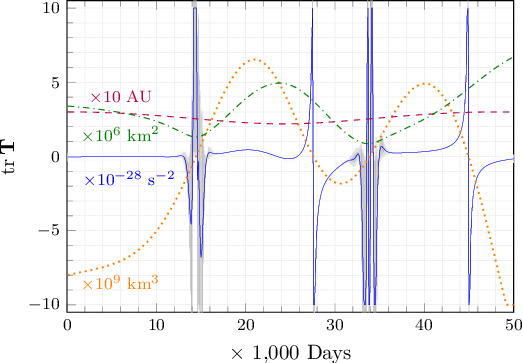}
\caption{\label{fig:str30AU}Simulation of a $\sim 30$~AU orbit, showing even greater sensitivity. Once numerical artifacts at standard double precision (left) are eliminated using 32-digit floating point arithmetic (right), sensitivities much better than $10^{-28}~{\rm s}^{-2}$ can be achieved throughout much of a $\sim$150 year orbit. Colors and line types are the same as in Fig.~\ref{fig:str1AU}.}
\end{figure}

\begin{figure}
\includegraphics[scale=0.8]{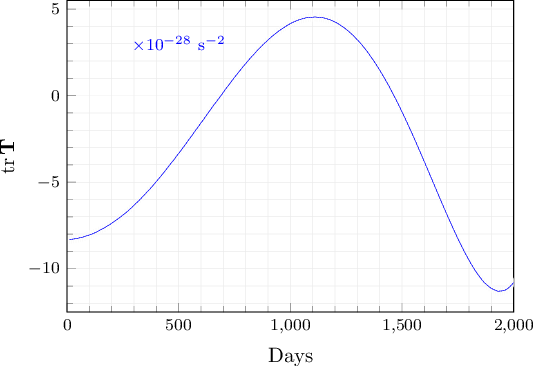}~\includegraphics[scale=0.8]{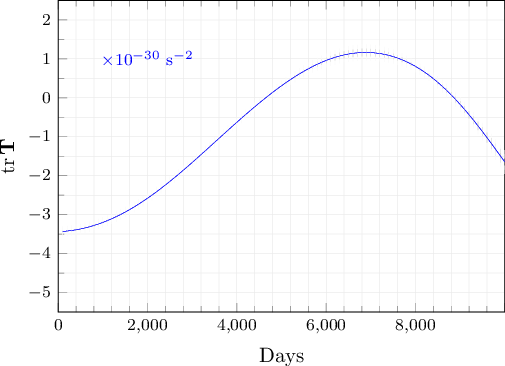}
\caption{\label{fig:ranges}Both the 10~AU (left) and the 30~AU (right) orbits offer exceptional sensitivity initially, while the tetrahedral configuration remains regular. (Later, as the orbits drift, the tetrahedron becomes increasingly distorted.) The 30~AU orbit, in particular, approaches the sensitivity required to detect the presence of the interplanetary matter density.}
\end{figure}

Going to even greater distances, at $\sim 30$~AU (Fig.~\ref{fig:str30AU}), the sensitivity of the configuration begins to approach the level that would be required, in principle, to detect the gravitational contribution of the interplanetary medium to $\nabla^2 U$. In particular, during the initial segment of their orbits, while the tetrahedron shape remains relatively free of distortion, both the 10~AU and the 30~AU configurations show exceptional sensitivity (Fig.~\ref{fig:ranges}).

We note that all these results, including the remarkable stability and accuracy of the $30$~AU orbital simulation, were achieved without accounting for angular accelerations, any linear acceleration of the tetrahedral configuration, or relativistic effects. This is again consistent with our earlier analytical estimate of uncertainties and errors, which showed that these contributions can be safely neglected. Ultimately, the major sources of error include the omission of angular acceleration and any nonlinear terms governing acceleration along the line of sight between satellites. These values represent distinct degrees of freedom that cannot be reconstructed from the observables that are available to us, namely a time series of intersatellite ranges and Sagnac-type timings.

As mentioned, to implement these theoretical sensitivities, instrumentation of exceptional accuracy may be required on board. Given a typical baseline of $10^6$~m between satellites, to achieve $\delta\tr\vec{T}\sim 10^{-23}~{\rm s}^{-2}$, it is necessary to measure intersatellite ranges at an accuracy better than ${\cal O}(10^{-7}~{\rm m})$, whereas acceleration estimates must be accurate to ${\cal O}(10^{-17}~{\rm m}/{\rm s}^2)$, which implies required clock accuracies better than ${\cal O}(10^{-15}~{\rm s})$. The satellites must, as a matter of course, also be ``clean'', free of any sources, internal or external, of nongravitational accelerations, including drag, solar radiation pressure, anisotropic thermal radiation from the vehicle, propellant or coolant leaks, or outgassing. Whether or not such stringent technical requirements are feasible is a question beyond the scope of our present study.

\section{Beyond Newton}
\label{sec:postnewton}

The tetrahedral constellation that we investigate is of particular interest as it may have the sensitivity to detect physics beyond Newtonian gravity. To this end, we made a first attempt to incorporate into our model various popular modifications of Newton's law of gravitation, to test by way of numerical experiment how these injected signals may show up in the reconstructed observable $\tr\vec{T}$.

For this investigation, we used the $\sim 10$~AU orbit of moderate eccentricity, shown in Fig.~\ref{fig:str10AU}. This choice is motivated by several factors:
\begin{itemize}
\item The sensitivity of the constellation increases at greater heliocentric distances.
\item The dependance on orbital radius varies between theories: e.g., the effects of MOND become more pronounced at greater distances from the Sun, while galileonic contributions diminish with distance.
\item An eccentric orbit may offer characteristic signatures that may be absent or harder to detect in more circular orbits.
\item A mission to greater heliocentric distances induces technical challenges related to its duration, communication, and power.
\end{itemize}

The orbit that we picked offers a suitable compromise addressing these competing factors, allowing us to contrast a baseline orbit against different paradigms of modified gravity. Specifically, we investigated
\begin{inparaenum}[a)]
\item a Yukawa term that corresponds to an interaction with a finite range, carried by a massive boson;
\item modification of the Newtonian law of gravitation by a short-range contribution due to a cubic galileon; and
\item modified Newtonian dynamics (MOND) and its acceleration scale dependent contribution to the Newtonian acceleration term.
\end{inparaenum}

\begin{figure}
\begin{center}
\includegraphics[scale=0.8]{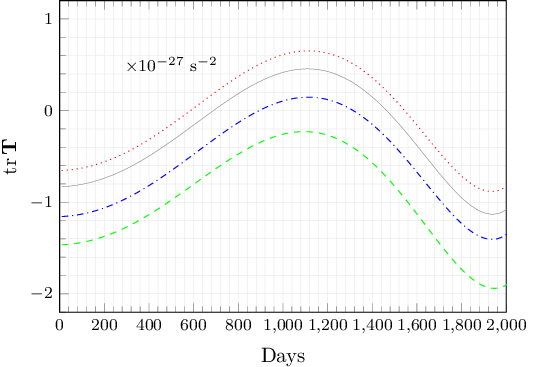}
\end{center}
\caption{\label{fig:plotsNew}Testing new physics during the first 2000 days of a constellation in a 10~AU orbit. In addition to the baseline measurement of $\tr\vec{T}$ (solid gray line), galileon gravity (green, dashed line), MOND (blue, dash-dotted line) and Yukawa gravity (red, dotted) are shown.}
\end{figure}

Moreover, the orbit we found has the advantage that initially, the tetrahedral configuration remains very stable, with exceptional sensitivity. This changes only after about 2000 days in orbit, at which time the tetrahedron becomes more degenerate, its volume collapsing and its baselines becoming more uneven as individual satellites follow individual orbits around the Sun.

Of course the example orbit that we used is not intended as a final or even as a recommended configuration. It simply serves as an example that demonstrates that a tetrahedral configuration may have sufficient sensitivity to detect ``new physics'' here in the solar system or, at the very least, establish new constraints on the parameters characterizing modified theories of gravitation.

%

\subsection{Yukawa contributions}

Modified theories of gravitation may involve fields with a nonzero rest mass and consequently, finite range. At the Newtonian level of approximation, such theories are characterized by a Yukawa-type modification of Newton's gravitational potential \cite{Adelberger2003}. We represent the Yukawa-potential, its gradient (the acceleration) and the trace of the gravitational gradient tensor as follows:
\begin{align}
\Phi&{}=-\frac{GM_\odot}{r}\big[1+\alpha(1-e^{-\mu r})\big],\\
\vec{a}=\nabla\Phi&{}=-\frac{GM_\odot}{r^3}\big[1 + \alpha\big(1 - (1 + \mu r)e^{-\mu r}\big)\big]\vec{r},\\
\nabla\cdot\vec{a}=\nabla^2\Phi&{}=-\alpha\mu^2\frac{GM}{r}e^{-\mu r}.\label{eq:YukawaT}
\end{align}
This representation, though less common, is equivalent to the more frequently seen form, $\Phi=-(GM_\odot/r)(1+\alpha e^{-\mu r})$, after the substitutions $G\Leftrightarrow (1+\alpha)G$, $\alpha\Leftrightarrow -\alpha/(1+\alpha)$. The representation we use has the advantage that in the $r\to 0$ limit it reduces to Newton's law of gravitation with $G$ being Newton's constant.

As an example, we considered the values $\alpha=10$, $\mu=(1~{\rm kpc})^{-1}$. With these values, using (\ref{eq:YukawaT}) we can estimate that the contribution to $\nabla\cdot\vec{a}$ will be $\sim 10^{-28}~{\rm s}^{-2}$, which is small but potentially detectable.

Our numerical simulation confirms this. As seen in Fig.~\ref{fig:plotsNew}, the Yukawa-contribution introduces a pronounced shift of $\tr\vec{T}$. One possible concern is that the shift being uniform, it may be difficult to distinguish a Yukawa-type contribution from baseline Newtonian gravity.

\subsection{Cubic galileons}

Galileon theories are scalar theories that introduce a scalar field invariant under the Galilean group. The simplest non-trivial extension beyond the quadratic theory involves cubic terms in the galileon Lagrangian \cite{Aslmarand2021}. The theory is screened at short range by the Vainshtein mechanism, characterized by the Vainshtein radius $r_V$. At short range, $r\ll r_V$, the theory yields the following acceleration law \cite{White2020}:
\begin{align}
\vec{a}&{}=-\frac{GM_\odot}{r^3}\left[1+\left(\frac{r}{r_0}\right)^{3/2}\right]\vec{r},\\
\nabla\cdot\vec{a}&{}=-\frac{3GM_\odot}{2(rr_0)^{3/2}}.\label{eq:GalT}
\end{align}
where we used the shorthand $r_0=(3/4)^{2/3}r_V$. We investigated the case characterized by $r_0=1$~kpc, for which we estimate a contribution of $\sim 6\times 10^{-28}~{\rm s}^{-2}$ to the trace of the gravitational gradient tensor (\ref{eq:GalT}) in the case of our selected orbit. As in the case of Yukawa gravity, Fig.~\ref{fig:plotsNew} demonstrates that a galileonic contribution shifts the observational value of $\tr\vec{T}$, as expected. Again, the uniform nature of the shift may present an observational challenge, but the magnitude of the shift demonstrates the sensitivity of the tetrahedral configuration.

%

%

\subsection{Modified Newtonian dynamics (MOND)}

Modified Newtonian Dynamics \cite{Milgrom1983}, or MOND, though it is merely an {\em ad hoc} modification of Newton's second law and not a bona fide modified theory of gravity, nonetheless remains perhaps the most popular method of modifying Newtonian gravity. The MOND version of Newton's force law, $\vec{F}=m\vec{a}\mu(a/a_0)$, includes an interpolating function that is characterized by the limits $\mu(a/a_0)\to 1$ for $a\gg a_0$, and $\mu(a/a_0)\to a_0/a$ for $a\lesssim a_0$, where $a_0\sim 1.2\times 10^{-10}~{\rm m}/{\rm s}^2$ is the MOND acceleration scale. The standard choice for the interpolating function is $\mu(x)=\sqrt{1/(1+x^2)}$, which leads to the following expressions for the gravitational acceleration and the trace of the gravitational gradient tensor in the gravitational field of the Sun:
\begin{align}
\vec{a}&{}=-\frac{GM_\odot}{r^3}\sqrt{\dfrac{1}{2}+\sqrt{\dfrac{1}{4}+\dfrac{a_0^2r^4}{G^2M_\odot^2}}}~\vec{r},\\
\nabla\cdot\vec{a}&{}=\frac{2\sqrt{2GM_\odot}~a_0^2r}{\sqrt{4a_0^2r^4+G^2M_\odot^2}\sqrt{\sqrt{4a_0^2r^4+G^2M_\odot^2}+GM_\odot}}.\label{eq:MONDT}
\end{align}
At $r\sim 10$~AU, the expression (\ref{eq:MONDT}) yields a contribution of $\sim 3\times 10^{-28}~{\rm s}^{-2}$ to the trace of the gravitational gradient tensor. This is confirmed by our simulation, as shown in Fig.~\ref{fig:plotsNew}.


\subsection{Relativistic corrections}

To estimate the extent to which general relativistic contributions affect our results, we turn to the post-Newtonian formalism. Acceleration due to a central compact mass $M$, such as the Sun, in a coordinate reference frame affixed to the source, would be given by \cite{MOYER2000}
\begin{align}
\vec{a}=-\frac{GM}{r^3}\vec{r}+\frac{GM}{c^2r^3}\left[
\left(\frac{2(\beta+\gamma)GM}{r}-\gamma v^2\right)\vec{r}+2(1+\gamma)(\vec{r}.\vec{v})\vec{v}\right],
\label{eq:relcorr}
\end{align}
where $\beta$ and $\gamma$ are Eddington's post-Newtonian parameters characterizing contributions by nonlinearity and spatial curvature. (For general relativity, $\beta=\gamma=1$.)

This formulation, in addition to terms with a nonzero contribution to $\nabla\cdot\vec{a}$, also includes velocity-dependent terms for which the concept of gradient is not readily applicable (i.e., there is no ``velocity field''). Moreover, including general relativistic corrections would also necessitate a complete revision of the method that we employed to convert to a satellite-fixed, noninertial coordinate system. It was for these reasons that we opted not to incorporate relativistic corrections into the model in the first place. Nonetheless, we can use (\ref{eq:relcorr}) to estimate the magnitude of the error that arises as a result of omitting these corrections:
\begin{align}
\delta_{\rm GR}\tr\vec{T}={\cal O}\left(\frac{(GM)^2}{c^2r^4}\right).
\end{align}
For a $r\sim 1$~AU orbit, the magnitude of the error is $\sim 10^{-21}~{\rm s}^{-2}$. This magnitude decreases by the fourth power of the configuration's heliocentric distance, becoming $\lesssim 10^{-27}~{\rm s}^{-2}$ in the case of a 30~AU orbit.

\subsection{Discussion}

The results presented in this section indicate that a tetrahedral configuration of satellites placed in a $\sim 10$~AU orbit has the sensitivity in principle to detect small deviations from Newtonian gravity, including a Yukawa force, cubic galileons or Modified Newtonian Dynamics. At the same time, the nature of the contribution of all three cases that we investigated is in the form of a near-constant shift in the value of $\tr\vec{T}$, which will likely pose a challenge when it comes to distinguishing a signal from the baseline Newtonian case. Moreover, the sensitivity is ultimately limited by the Newtonian formalism, which forms the conceptual basis for this experiment.

What we nonetheless demonstrated through these calculations and simulation results is that the signals are unambiguous, and remain present in the reconstructed value of $\tr\vec{T}$ that, in our simulation, was computed using only observables that will be available during such a mission: notably, intersatellite ranges, the Sagnac-type observable used to ``derotate'' the satellite-fixed reference frame, and the approximate direction and distance to the Sun in the satellite-fixed reference frame.

As we emphasized, the orbits chosen for this simulation wer picked for demonstration only. It is not suggested by any means that our choices are optimal. Indeed, should such a tetrahedral configuration be seriously considered as an actual mission, finding an optimal orbit will be one of the major challenges. The choice will have to balance between the stability of the tetrahedral configuration as the satellites progress along their individual trajectories, the variable distance from the Sun and its impact, both positive and negative, on the detectability of any signal and the ability to distinguish it from the baseline case, and of course, the technical challenges involved, which we do not discuss in the present study.

\section{Conclusions}
\label{sec:summary}

In this study, we revisited the concept of using a tetrahedral constellation of satellites to carry out highly accurate measurements of the trace of the gravitational gradient tensor in interplanetary space.

When we first explored this topic in \cite{TOTH2023ApSS}, we obtained analytical estimates of the sensitivity of the constellation, but due to the technical limitations of standard machine-precision arithmetic, we were only able to offer limited support by way of computer simulations. This has now been remedied: our more recent simulations were carried out at extended precision using an appropriate numerical library. In the process, we also addressed shortcomings in our original software code, and introduced improved algorithms for numerical differentiation and integration.

As a result of these efforts, a remarkable picture emerges. First, we confirmed that for heliocentric orbits in the vicinity of the Earth, our earlier analytical estimate of the sensitivity of the constellation appear valid: the gravitational gradient tensor may be measurable to the tune of ${\cal O}(10^{-23}~{\rm s}^{-2})$. This sensitivity might be sufficient to validate or exclude certain families of modified gravity theories although ultimately limited by the formalism that ignores relativistic corrections.

We also explored the solution space by considering larger orbits. Though such missions may be technically more challenging and may take longer to carry out, they may offer unprecedented opportunities to test the theory of gravitation in the solar system. An orbit with a semi-major axis of $\sim$10~AU might reach sensitivities of ${\cal O}(10^{-28}~{\rm s}^{-2})$ during certain orbital segments, whereas a 30~AU orbit can reach an absolutely exceptional sensitivity of ${\cal O}(10^{-30}~{\rm s}^{-2})$ under ideal circumstances. This value, in particular, places within reach experiments that might detect the gravitational contribution of the interplanetary medium, and ultimately, perhaps even contributions from dark matter and dark energy. Conversely, such measurements may offer exceptionally sensitive tests of modified gravity here in the solar system, especially theories that predict deviations from the vacuum Poisson equation, which includes families of theories that yield a MOND-like modified law of inertia.

To this end, we explored three specific scenarios involving three popular modifications of Newton's law of gravitation: Yukawa gravity, a galileonic field, and last but not least, Modified Newtonian Dynamics. In all three cases we found that, using physically justifiable parameterizations of these modified theories, we obtain a signal of detectable magnitude using a tetrahedral configuration of satellites in a 10~AU orbit. In addition to estimating the magnitude of the contribution to the trace of the gravitational gradient tensor from theory, we performed a simulation that, we stress, modeled the actual observable: the trace reconstructed from simulated measurements of intersatellite range observables, Sagnac-type observables measuring the rotation of the configuration, and the approximate distance from, and direction to, the Sun.

Should such a constellation be considered, however, there remain numerous challenges. As we described, though the signal is clearly present at detectable levels, it may be difficult to distinguish it from the baseline value that would be due solely to Newtonian gravity. The technical challenges are also formidable, requiring exceptional accuracy when measuring intersatellite ranges or the timing and interferometry required for the Sagnac-type observable. Conceivably, an experiment may benefit from the lessons already learned from, or lessons that may be offered by, experiments like the GRACE follow-on \cite{REL2014} or LISA/LISA Pathfinder \cite{LISA2013} missions.

Rather than achieving marginal improvements, e.g., by introducing additional satellites, one possible strategy may instead rely on multiple tetrahedral configurations in orbits with different semimajor axes and eccentricities, perhaps even different inclinations. Such a system may help map deviations from Newtonian gravity in the three-dimensional domain of the solar system, allowing us to unambiguously distinguish Newtonian gravity from any non-Newtonian contributions, if present.

As a first step in this direction, however, our current study and simulation demonstrates that in-situ measurements of the trace of the gravitational gradient tensor in the solar system, at heliocentric distances and timescales that are within the realm of the feasible, are possible in principle.

\section*{Acknowledgments}
VTT thanks Slava Turyshev for discussions and acknowledges the generous support of Vladimir Andreev, Plamen Vasilev and other Patreon patrons.

\section*{References}
\bibliography{refs}

\begin{thebibliography}{10}

\bibitem{TOTH2023ApSS}
Viktor~T. {Toth}.
\newblock {Gravitational anomaly detection using a satellite constellation:
  analysis and simulation}.
\newblock {\em \apss}, 368(10):92, October 2023.

\bibitem{Yu2019}
Nan Yu, Sheng-wey Chiow, Jerome Gleyzes, Phil Bull, Olivier Dore, Jason Rhodes,
  Jeffrey Jewell, Eric Huff, and Holger Muller.
\newblock {Direct Probe of Dark Energy Interactions with a Solar System
  Laboratory}, 2019.

\bibitem{Nicolis2009}
Alberto {Nicolis}, Riccardo {Rattazzi}, and Enrico {Trincherini}.
\newblock {Galileon as a local modification of gravity}.
\newblock {\em \prd}, 79(6):064036, March 2009.

\bibitem{Turyshev2024}
Slava~G. {Turyshev}, Sheng-wey {Chiow}, and Nan {Yu}.
\newblock {Searching for new physics in the Solar System with tetrahedral
  spacecraft formations}.
\newblock {\em \prd}, 109(8):084059, April 2024.

\bibitem{REL2014}
Slava~G. Turyshev, Mikhail~V. Sazhin, and Viktor~T. Toth.
\newblock {General relativistic laser interferometric observables of the
  GRACE-Follow-On mission}.
\newblock {\em Phys. Rev. D}, 89:105029, May 2014.

\bibitem{LISA2013}
Paul~W. {McNamara}.
\newblock {The LISA Pathfinder Mission}.
\newblock {\em International Journal of Modern Physics D}, 22(1):1341001,
  January 2013.

\bibitem{Sagnac1}
Georges {Sagnac}.
\newblock L'{\'e}ther lumineux d{\'e}montr{\'e} par l'effet du vent relatif
  d'{\'e}ther dans un interf{\'e}rom{\`e}tre en rotation uniforme.
\newblock {\em Comptes Rendus}, 157:708--710, 1913.

\bibitem{Sagnac2}
Georges {Sagnac}.
\newblock Sur la preuve de la r{\'e}alit{\'e} de l'{\'e}ther lumineux par
  l'exp{\'e}rience de l'interf{\'e}rographe tournant.
\newblock {\em Comptes Rendus}, 157:1410--1413, 1913.

\bibitem{Rodrigues1840}
Rodrigues.
\newblock Des lois g{\'e}om{\'e}triques qui r{\'e}gissent les d{\'e}placements
  d'un syst{\`e}me solide dans l'espace, et de la variation des coordonn{\'e}es
  provenant de ces d{\'e}placements consid{\'e}r{\'e}s ind{\'e}pendamment des
  causes qui peuvent les produire.
\newblock {\em Journal de Math{\'e}matiques Pures et Appliqu{\'e}es}, pages
  380--440, 1840.

\bibitem{decimaljs}
Michael McLaughlin.
\newblock decimal.js: An arbitrary-precision decimal type for javascript.
\newblock \url{https://github.com/MikeMcl/decimal.js/}, 2021.
\newblock Accessed: 2024-05-19.

\bibitem{Runge1895}
C.~Runge.
\newblock Ueber die numerische auflösung von differentialgleichungen.
\newblock {\em Mathematische Annalen}, 46:167--178, 1895.

\bibitem{Kutta1901}
W.~Kutta.
\newblock Beitrag zur näherungsweisen integration totaler
  differentialgleichungen.
\newblock {\em Zeitschrift Math. Phys.}, 46:435--453, 1901.

\bibitem{Yoshida1990}
Haruo {Yoshida}.
\newblock {Construction of higher order symplectic integrators}.
\newblock {\em Physics Letters A}, 150(5-7):262--268, November 1990.

\bibitem{stormer1907}
Carl St{\"o}rmer.
\newblock {Sur les trajectoires des corpuscules {\'e}lectris{\'e}s dans
  l'espace. Applications {\`a} l'aurore bor{\'e}ale et aux perturbations
  magn{\'e}tiques}.
\newblock {\em {Radium (Paris)}}, 4(1):2--5, 1907.

\bibitem{Richardson1911}
L.~F. {Richardson}.
\newblock {The Approximate Arithmetical Solution by Finite Differences of
  Physical Problems Involving Differential Equations, with an Application to
  the Stresses in a Masonry Dam}.
\newblock {\em Philosophical Transactions of the Royal Society of London Series
  A}, 210:307--357, January 1911.

\bibitem{TOTH2008}
Viktor~T. Toth.
\newblock {Independent analysis of the orbits of Pioneer 10 and 11}.
\newblock {\em International Journal of Modern Physics D}, 18:717, 2009.

\bibitem{Adelberger2003}
E.~G. {Adelberger}, B.~R. {Heckel}, and A.~E. {Nelson}.
\newblock {Tests of the Gravitational Inverse-Square Law}.
\newblock {\em Annual Review of Nuclear and Particle Science}, 53:77--121,
  December 2003.

\bibitem{Aslmarand2021}
Shahabeddin~M. {Aslmarand}, Amin~Rezaei {Akbarieh}, Yousef {Izadi}, Sobhan
  {Kazempour}, and Lijing {Shao}.
\newblock {Cosmological aspects of cubic Galileon massive gravity}.
\newblock {\em \prd}, 104(8):083543, October 2021.

\bibitem{White2020}
Nicholas~C. {White}, Sandra~M. {Troian}, Jeffrey~B. {Jewell}, Curt~J. {Cutler},
  Sheng-wey {Chiow}, and Nan {Yu}.
\newblock {Robust numerical computation of the 3D scalar potential field of the
  cubic Galileon gravity model at solar system scales}.
\newblock {\em \prd}, 102(2):024033, July 2020.

\bibitem{Milgrom1983}
M.~{Milgrom}.
\newblock {A modification of the Newtonian dynamics as a possible alternative
  to the hidden mass hypothesis}.
\newblock {\em \apj}, 270:365--370, July 1983.

\bibitem{MOYER2000}
Theodore~D. Moyer.
\newblock {\em Formulation for Observed and Computed Values of Deep Space
  Network Data Types for Navigation}.
\newblock John Wiley \& Sons, 2005.

\end{thebibliography}
\bibliographystyle{unsrt}

\appendix

\renewcommand{\theequation}{A\arabic{equation}}
\setcounter{equation}{0}

\end{document}